\documentclass[twocolumn,aps,pra,floatfix,footinbib,notitlepage,superscriptaddress,groupaddress,showpacs,showkeys]{revtex4-1}
\usepackage{graphicx,graphics,times,bm,bbm,bbold,amssymb,amsmath,amsfonts,dsfont,hyperref,mathrsfs,color,caption,subcaption,cancel}
\usepackage{upgreek}
\usepackage[utf8]{inputenc}
\usepackage{amsmath}
\usepackage{xcolor}
\usepackage{mathtools,amssymb}
\usepackage{amsmath, nccmath}
\usepackage{lipsum}
\usepackage{blindtext}
\usepackage{graphicx}
\usepackage{dcolumn}
\usepackage{hyperref}
\hypersetup{
    colorlinks,
    linkcolor={blue},
    citecolor={blue},
    urlcolor= {blue}
}

\newcommand{\ignore}[1]{}

\let\oldsqrt\sqrt
\def\sqrt{\mathpalette\DHLhksqrt}
\def\DHLhksqrt#1#2{%
\setbox0=\hbox{$#1\oldsqrt{#2\,}$}\dimen0=\ht0
\advance\dimen0-0.2\ht0
\setbox2=\hbox{\vrule height\ht0 depth -\dimen0}%
{\box0\lower0.4pt\box2}}

\DeclareFontFamily{OT1}{pzc}{}
\DeclareFontShape{OT1}{pzc}{m}{it}%
              {<-> s * [1.25] pzcmi7t}{}
\DeclareMathAlphabet{\mathpzc}{OT1}{pzc}%
                                 {m}{it}

\usepackage{amsmath}
\usepackage{etoolbox}

\makeatletter
\newcommand{\changeoperator}[1]{%
  \csletcs{#1@saved}{#1@}%
  \csdef{#1@}{\changed@operator{#1}}%
}
\newcommand{\changed@operator}[1]{%
  \mathop{%
    \mathchoice{\textstyle\csuse{#1@saved}}
               {\csuse{#1@saved}}
               {\csuse{#1@saved}}
               {\csuse{#1@saved}}%
  }%
}
\makeatother

\changeoperator{sum}
\changeoperator{prod}

\makeatletter
\renewcommand\@makecaption[2]{%
  \par
  \vskip\abovecaptionskip
  \begingroup
   \small\rmfamily
    \begingroup
     \samepage
     \flushing
     \let\footnote\@footnotemark@gobble
     \@make@capt@title{#1}{#2}\par
    \endgroup
  \endgroup
  \vskip\belowcaptionskip
}
\makeatother

\begin{document}


\title{NISQ-Compatible Error Correction of Quantum Data Using Modified Dissipative Quantum Neural Networks} 
\author{Armin Ahmadkhaniha}
 \email[ ]{ahmadkhaniha@ut.ac.ir}
\affiliation{School of Electrical and Computer Engineering, University of Tehran, Tehran, Iran }
\author{Marzieh Bathaee}
 \email[ ]{neda.bathaee@gmail.com}
\affiliation{ Sharif Quantum Center, Sharif University of Technology, Tehran 14588-89694, Iran }

\begin{abstract}
Using a dissipative quantum neural network (DQNN) accompanied by conjugate layers, we upgrade the performance of the existing quantum auto-encoder (QAE) network as a quantum denoiser of a noisy m-qubit GHZ state. Our new denoising architecture requires a much smaller number of learning parameters, which can decrease the training time, especially when a deep or stacked DQNN is needed to approach the highest fidelity in the Noisy Intermediate-Scale Quantum (NISQ) era. In QAE, we reduce the connection between the hidden layer's qubits and the output's qubits to modify the decoder. The Renyi entropy of the hidden and output qubits' states is analyzed with respect to other qubits during learning iterations. During the learning process, if the hidden layer remains connected to the input layers, the network can almost perfectly denoise unseen noisy data with a different underlying noise distribution using the learning parameters acquired from training data.

\end{abstract}
\maketitle

\section{Introduction}
\label{intro}
Artificial intelligence (AI) is almost increasingly ubiquitous in science and everyday life. Using machine learning (ML) techniques as a sub-field of AI to find the critical feature of big or complex data is the central part of today's industry and high-level scientific research such as biology, chemistry, and physics \cite{maheshwari2022quantum,orus2019tensor}. Employing Neural Networks in ML has had a tremendous victory in fields such as image processing, language recognition, and compressing and denoising data \cite{krizhevsky2012imagenet,radford2018improving}.

Unfortunately, Moore's law \cite{658762} is saturating while the amount of data for processing keeps skyrocketing. However, the quantum mechanical theory promises a fundamental solution in speeding up data processing. Technologies based on counterintuitive quantum properties, such as non-locality and superposition, manifest their extraordinary potential in higher-performance computations and information processing, totally different from the classical ones.  Quantum computers (QC) may be just around the corner. Utilizing this quantum power in machine learning tasks has introduced a new field dubbed quantum machine learning (QML) at the interface of AI and quantum information processing. QML manipulates quantum data via quantum algorithms run on parametrized quantum circuits \cite{schuld2021machine, Lamata_2020,biamonte2017quantum}. The quantum data is either embedded from classical datasets or generated in quantum experiments \cite{williams2023quantum}. 

Nevertheless, currently implemented quantum processors are in the Noisy Intermediate-Scale Quantum (NISQ). Thus, studying hardware-efficient quantum algorithms and their performance is increasing through deterministic quantum algorithms such as Grover \cite{ahmadkhaniha2023performance} as well as variational quantum algorithms \cite{cerezo2021variational}.
Therefore, due to the noise present in quantum data, performing a thorough characterization of the device and post-processing is necessary. The noise impacting quantum processors can be addressed by utilizing the power of QML techniques such as quantum auto-encoder (QAE).
In classical ML, an auto-encoder (AE) is a deep neural network (NN) with a bottleneck \cite{LIOU201484} utilized for compressing and denoising data \cite{Romero_2017,grant2018hierarchical,achache2020denoising}. The basic building block of NN is a perceptron that connects input neurons to output neurons through a learnable function \cite{goodfellow2016deep}. A quantum perceptron must be introduced when the input data is quantum data, and subsequently, the quantum neural network must be built. The simplest model for a quantum perceptron comprises unitary operators implemented by parametrized quantum gates, which evolve the initial state of the quantum input state to a final quantum state \cite{beer2020training}. Here, for instance, quantum states can be quantum bits (qubits) playing the roles of neurons in the classical counterpart.

Different configurations of single and two-qubit gates have been proposed and implemented for the quantum perceptrons of quantum neural networks (QNN) \cite{Beer2019EfficientLF}. 
Like conventional NN, any QNN is characterized by the number of layers and qubits in each layer. The former shows how the network is deep, and the latter indicates the width of the quantum network.
The latest and high-performed QNN is a dissipative quantum neural network (DQNN) where its NISQ-compatible version ($\text{DQNN}_{\text{NISQ}}$) was addressed in \cite{beer2022quantum}. The advantage distinguishing the DQNN architecture from the previous ones is that the total number of qubits scales with only the maximum width of QNN, not the depth of QNN \cite{beer2021training}. This superiority comes from the fact that in DQNN, only the output qubits of each layer contribute as input to the next layer via tracing out the other degrees of freedom related to the previous layer. 

For denoising the noisy quantum data such as an m-qubit GHZ state \cite{greenberger1989going}, a quantum autoencoder with a similar structure to AE has been studied vigorously \cite{doi:10.1126/sciadv.abn9783,Lamata_2019,Liu_2021}. The QAE implemented by QNN, similar to DQNN, has also been studied in \cite{PhysRevLett.124.130502}. The quantum stacked, deep QAEs, and a more general version of modified QAE named brain-box QAE (see Fig.~\ref{QAE}~(a)) for denoising the m-qubit GHZ states have been investigated \cite{PhysRevLett.124.130502,pazem2023error}.
Taking advantage of overparametrization \cite{larocca2023theory}, QAE with brain-box simulation in \cite{pazem2023error} showed higher noise tolerance. However, the NISQ-compatible quantum networks still demand variational quantum circuits with the minimum possible complexity. So, it is absolutely essential to both minimize the number of learning parameters and fabricate quantum neural networks with higher noise tolerance.
Like conventional NN, QNN encounters problems such as barren plateau, overfitting, and instability \cite{sharma2022trainability,williams2023quantum,PhysRevLett.128.180505}. Moreover, finding the best hyper-parameters, such as the learning rate, is still a critical issue in QML \cite{Borras_2023}. While in conventional ML, the dropout technique helps to eliminate overfitting, recently \cite{chalkiadakis2023exploring} illustrated using conjugate layers, one can decrease the learning parameters and avoid overfitting. The conjugate layers are randomly chosen from the QNN layers, and their quantum perceptrons are Hermitian conjugate quantum operators of one of the previous layers. Note that the layer and its conjugate have the same learning parameters. 

This paper investigates the NISQ-compatible DQNN ($\text{DQNN}_{\text{NISQ}}$) equipped without and with the conjugate layers in QAE to enhance the denoising procedure of m-qubit GHZ. We have modified the quantum network of the QAE decoder, and our experiments show that the revised QAE outperforms the original version. The second-order quantum Renyi entropy is derived to measure the purity of the QAE's output state during the learning process. We discover less direct connectivity between output qubits and the bottleneck's qubits; the quantum output state (denoised GHZ) becomes separable from other qubits of the quantum network, and the validation data is denoised better.
To clarify the importance of our work, we must emphasize that our trained QAE circuit has the advantage compared to the previous results \cite{PhysRevLett.124.130502,achache2020denoising,pazem2023error}  that it needs much fewer parameters, leading to a decrease in the cost of the quantum gate implementation.

The material of the paper is organized as follows. In section~\ref{DQNN+con}, we compare the results of denoising the noisy GHZ state in NISQ-compatible DQNN with and without conjugate layers. Section~\ref{Modified decoder} proposes modifying the decoder's quantum perceptrons to enhance the QAE performance. We also discuss the reasons behind the outperformance of our model by analyzing Renyi entropy as a figure of merit during the training. 
We validate our proposed idea using Pennylane \cite{bergholm2022pennylane}, and TensorFlow \cite{tensorflow2015-whitepaper}. The paper is concluded in section~\ref{con}.

\section{$\text{DQNN}_{\text{NISQ}}$-based QAE }
\label{DQNN+con}

This section first studies a QAE that employs the $\text{DQNN}_{\text{NISQ}}$. Hereafter, we call this structure $\text{QAE}_{\text{NISQ}}$. Then, the $\text{QAE}_{\text{NISQ}}$ is modified by a conjugate layer. The unsupervised learning process of QAE for denoising an m-qubit GHZ state is presented and the numerical results are compared.

\subsection{$\text{DQNN}_{\text{NISQ}}$-based QAE Structure}

\begin{figure}[htb]
\centering
\includegraphics[width=\linewidth]{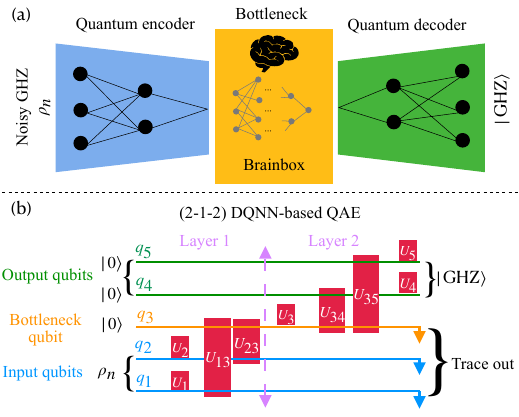}
\caption{(Color online) (a) Schematic configuration of a generic Quantum Auto Encoder. (b) (2-1-2) architecture of DQNN-based QAE. $q_i$ stands for the $i$th qubit.}
\label{QAE}
\end{figure}

As shown in Fig.~\ref{QAE}(a), the structure of a quantum auto-encoder, similar to the conventional one, comprises an encoder, a bottleneck (a brainbox), and a decoder. As an unsupervised machine learning, AE puts the essence of the input data set information (features) to the bottleneck and then reproduces input data in the decoder part. In the quantum version of AE, i.e., QAE, the encoder, brainbox, and decoder are parametrized quantum circuits that must be designed efficiently. 

In the NISQ era, empirically preparing a noiseless quantum state is impossible. Therefore, QAE is utilized for error correction of quantum data. One of the quantum resources to have highly secured networks is the Greenberger–Horne–Zeilinger state (GHZ).
An m-qubit GHZ is a highly entangled state which is presented in the computational basis $|0\rangle,|1\rangle$ as $|\text{GHZ}\rangle=\frac{1}{\sqrt{2}}(|00\cdots 0\rangle+|11\cdots 1\rangle)$, where $|0\rangle$ and $|1\rangle$ are repeated $m$ times in each ket. 
This paper uses a symmetric bit-flip noise to influence the m-qubit GHZ state as training data. At the same time, the denoising procedure is evaluated by quantum data affected by a quantum depolarizing channel. 

Figure~\ref{QAE}(b) depicts the DQNN-based QAE for denoising a 2-qubit GHZ state with one qubit in its bottleneck. The NISQ-compatible DQNN involves parametrized single-qubit and two-qubit gates, which guarantees the universality of the quantum circuit. The single-qubit gate $U_i,~i\in\{1,..,m\},$ has three learnable parameters. The canonical two-qubit gate $U_{ij},~i,j\in\{1,..,m\},$ comprises three two-qubit rotation gates, i.e., $R_{X_iX_j}$, $R_{Y_iY_j}$, and $R_{Z_iZ_j}$, where $R_{A_iA_j}=\exp(i\theta_A A_iA_j)$, and $A_i\in\{X_i,Y_i,Z_i\}$ is the Pauli matrix acting on the qubit $i$ \cite{zhang2003geometric}. Since each two-qubit rotation has a specific learnable parameter $\theta_A$, the two-qubit gate $U_{ij}$ is indicated by three learnable parameters. Hence, the ($m\text{-1-}m$)~$\text{QAE}_{\text{NISQ}}$
has $n_{par}=3(4m+1)$ training parameters indicated by $\vec\theta$ from now on. 

It is worth mentioning that \cite{beer2020training,PhysRevLett.124.130502,pazem2023error} investigated QAE equipped with DQNN while the unitary operators constructing the quantum perceptrons were any possible combinations of Pauli matrices. Hence, they have more learning parameters than the $\text{QAE}_{\text{NISQ}}$ proposed in this paper.

\subsection{The training strategy of QAE }
The quantum data set of QML is divided into two categories, i.e.,  training and validation data set. 
Unsupervised machine learning is the driving force behind the QAE, so the dataset lacks labels. 
When QAE performs as a denoiser, its output must be assessed by the perfect input version.  In our case, the noisy GHZ must be evaluated by a perfect pure GHZ state. However, there is no such perfect reference in reality. Moreover, the no-cloning theorem prevents preparing a copy of a perfect GHZ state for training. Therefore, the output of the QAE fed by $i$th training data, i.e., bit-flip noisy GHZ states $|\text{GHZ}\rangle_{t_i,p}$ with the probability $p$, must be compared to the other bit-flip noisy GHZ state as a reference $|\text{GHZ}\rangle_{r_i,p}$. This comparison will be performed by calculating the fidelity between them as 
\begin{eqnarray}
F_i(\vec\theta)=_{r_i,p}\langle\text{GHZ}|\rho_\text{GHZ}^{t_i,p}(\vec\theta)|\text{GHZ}\rangle_{r_i,p}, 
\label{eq-fidelity-0}
\end{eqnarray}
where
\begin{eqnarray}
\rho_\text{GHZ}^{t_i,p}(\vec\theta)=\text{Tr}_{q}(U(\vec\theta)|\text{GHZ}\rangle_{t_i,p}\langle\text{GHZ}|\otimes(|0\rangle\langle0|)^{\otimes |q|}U^{\dagger}(\vec\theta)).
\nonumber\\
\label{eq-fidelity}
\end{eqnarray}
In Eq.~(\ref{eq-fidelity}), $U(\vec\theta)$ is the unitary operator describing the DQNN-based QAE's circuit evolution. Moreover, $q$ is the set of qubits traced out at the output layers, and $|q|$ is the number of such qubits.
For instance, for the (2-1-2) QAE presented in Fig.~\ref{QAE}, $U(\vec\theta)=U_5U_4U_{35}U_{34}U_3U_{23}U_{13}U_2U_1$ and $q=\{q_1,q_2,q_3\}$. Recall that each single-qubit gate $U_i$ and two-qubit gate $U_{ij}$ have three learning parameters in the structure of $\text{DQNN}_{\text{NISQ}}$. 
Moreover, an ancilla-free swap test measures the fidelity of two states, as described in \cite{PhysRevA.87.052330}.
This paper assumes the training data is symmetrically bit-flipped with the same probability $p$ on each qubit. In contrast, the quantum depolarizing channel (QDC) affects the validation data with different probabilities on each qubit. Hence, the QAE denoises the bit-flip noisy data and evaluates its performance with the validation data with another noise distribution.

Like ML, a cost function is introduced and optimized to learn the best parameters of the quantum neural network's quantum circuits, which leads to denoising quantum data. Accordingly, one can follow one of the two approaches for defining the cost function for the QAE neural network. 

\begin{itemize}
\item The average-based cost function is formed by averaging the fidelity over the whole elements of a fixed training data set, i.e., $C_{av}(\vec\theta)=\frac{1}{N}\sum_{i=1}^NF_i(\vec\theta)$, where $N$ is the number of the training data set. Note that this average is classical and different from the quantum expectation that must be applied to calculate the fidelity \cite{PhysRevA.105.062404}. 
At each iteration of the quantum neural network's training, the learning parameters $\vec\theta$ are updated based on a predetermined optimization on the cost function involving the whole training data set, such as SGD or ADAM \cite{jordan2015machine}. As a result, for the $N$ training data, the average cost function is calculated in each iteration, and accordingly, based on the opted optimization, the learning parameters are updated.

\item The individual data-based cost function is defined by the fidelity of the $i$th training data, i.e., $C_i(\vec\theta)=F_i(\vec\theta)$ and the learning parameters are updated based on individual training data. In fact, this strategy uses the stochastic gradient descent idea in which learning parameters are updated based on one sample per step. Since each data is randomly chosen from the bit-flipped noisy distribution, the number of iterations of DQNN equals the cardinality of the training data set, i.e., $N$.  

\end{itemize}

It is noteworthy to state that in real quantum hardware, various methods exist to implement the fidelity gradient, such as parameter shift \cite{PhysRevA.98.032309} or quantum gradients \cite{PhysRevA.98.012324}.

In our simulation, the learning process with the individual data-based cost function is much faster than the average-based one. However, it requires a higher number of training data. Since the data-based cost function assesses only one training data, there is a high chance of inaccurate training, i.e., the learned quantum circuit is not verified via validation data. Therefore, this paper only uses this method to randomly initialize the learning parameters around the parameters learned by the individual data-based cost function.

\subsection{$\text{DQNN}_{\text{NISQ}}$-based QAE with a Conjugate Layer}

\begin{figure}[htb]
\centering
\includegraphics[width=\linewidth]{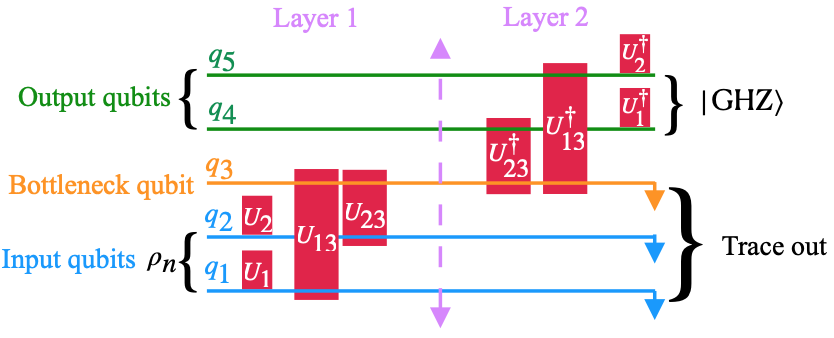}
\caption{(Color online) Schematic configuration of a (2-1-2)~architecture of $\text{QAE}_{\text{NISQ}}$ with a conjugate layer.}
\label{modified2-1-2}
\end{figure}

Like the dropout and U-net techniques in ML, reducing the learning parameters may alleviate the challenges such as training plateau and training time; conjugate layers in QNN serve an akin role \cite{chalkiadakis2023exploring}. As mentioned before, the unitary operator indicates each layer in QNN. Therefore, a unitary operator of a conjugate layer is determined by the time-reversal unitary operator corresponding to one of the previous layers of QNN or the Hermitian conjugate operator of it. Therefore, since the learning parameters of the layer and its conjugate layer are the same, the total number of learning parameters decreases. 

As depicted in Fig.~\ref{QAE}, a quantum auto-encoder encodes data to its compressed form and decodes it to its original. In other words, an input quantum state evolves via a unitary operator in encoding circuits, and its Hermitian conjugate unitary operator must be applied afterward in the decoder to give the same input quantum state. Thus, the idea of using conjugate layers in the
$\text{QAE}_{\text{NISQ}}$ can be applicable. To do this, as Fig.~\ref{modified2-1-2} shows, in the $\text{QAE}_{\text{NISQ}}$ equipped with conjugate layers, henceforth we call it $\text{QAE}_{\text{conj}}$, the quantum gates of layer 2 are the Hermitian conjugate of the quantum gates of layer 1. As a result, the learning parameter becomes half compared to the $\text{DQNN}_{\text{NISQ}}$-based QAE without conjugate layers ($\text{QAE}_{\text{NISQ}}$). For ($m\text{-1-}m$)~$\text{QAE}_{\text{conj}}$, the learning parameter is $\bar{n}_{par}=6m$.

\subsection{Simulation}

\begin{figure}[htb]
\centering
\includegraphics[width=\linewidth]{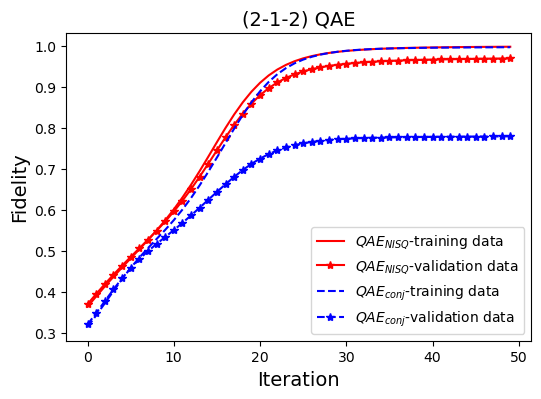}
\caption{(Color online) Comparison of the fidelity change during a (2-1-2)~QAE training process based on DQNN with (dashed blue) and without (solid red) conjugate layers. The starred lines relate to the validation data applied to the learned quantum circuit during the iteration.
The bit-flip noise probability is $p=0.2$. The number of training data is 30. The optimization method is stochastic gradient descent (SGD) with a learning rate of 0.4, and the average cost function is used.}
\label{conj-result}
\end{figure}

Figure~\ref{conj-result} illustrates the simulation of fidelity change during the denoising procedure of the (2-1-2) QAE when the average cost function is used for 30 training data. In both quantum networks, $\text{DQNN}_{\text{NISQ}}$-based QAE without and with the conjugate layer, the validation data does not coincide with the training data at the end of the training. Although the number of learning parameters in $\text{QAE}_{\text{conj}}$ is half of the $\text{QAE}_{\text{NISQ}}$, the quantum network's performance is inferior. This can confirm that the over-parametrization technique \cite{larocca2023theory} improves the QNN performance.  In the next section, we describe a modification of the decoder part of $\text{QAE}_{\text{conj}}$. Then, we will show the superiority of the $\text{DQNN}_{\text{NISQ}}$-based QAE with a conjugate layer and a modified decoder.  We call this new structure $\text{QAE}_{\text{conj+mod~dec}}$ from now on.  During learning, the Renyi entropy change of the hidden layer can determine which quantum perceptron structures exhibit superiority.

\begin{figure}[htb]
\centering
\includegraphics[width=\linewidth]{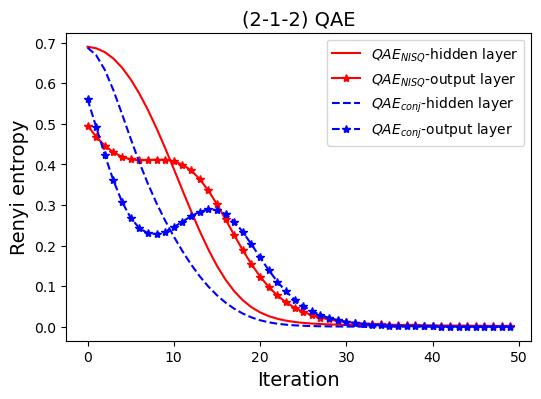}
\caption{(Color online) Comparison of the Renyi entropy change of outputs and hidden layer (bottleneck) during a (2-1-2) QAE training process based on $\text{DQNN}_{\text{NISQ}}$ with (dashed blue) and without (solid red) conjugate layers. The starred lines relate to the output qubits Renyi entropy during the iteration. The bit-flip noise probability is $p=0.2$. The number of training data is 30. The optimization method is stochastic gradient descent (SGD) with a learning rate of 0.4.}
\label{renyi-conj}
\end{figure}

The second order, Renyi entropy $E_R=-\ln (\text{Tr}\rho^2)$, is a parameter that quantifies the purity of a quantum state $\rho$. Figure~\ref{renyi-conj} shows the variation of the Renyi entropy of the state of the bottleneck qubit and the output qubits for the (2-1-2)~QAE. As expected, when the fidelity reaches 1, the state of the output qubits becomes separable. For $\text{QAE}_{\text{NISQ}}$ and $\text{QAE}_{\text{conj}}$, the bottleneck qubit is also separable after the fidelity of the output is set to one. So, the correlation between the hidden layer and other layers decreases to zero during the learning. 

\begin{figure}[htb]
\centering
\includegraphics[width=\linewidth]{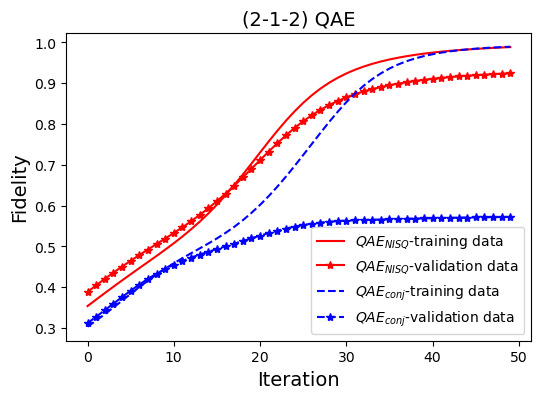}
\caption{(Color online) Comparison of the fidelity change during a (2-1-2) QAE training process based on $\text{DQNN}_{\text{NISQ}}$ without (solid red) and with (dashed blue) a conjugate layer. The starred lines relate to the validation data applied to the learned quantum circuit during the iteration. The bit-flip noise probability is $p=0.4$. The number of training data is 30. The optimization method is stochastic gradient descent (SGD) with a learning rate of 0.4, and the average cost function is used.}
\label{212p04fid}
\end{figure}

Another important feature that must be evaluated for the $\text{DQNN}_{\text{NISQ}}$-based QAE as a denoiser is its noise tolerance. Figures \ref{212p04fid} illustrate the fidelity of the $\text{QAE}_{\text{NISQ}}$ and $\text{QAE}_{\text{conj}}$ with noise threshold $p=0.4$, respectively. Comparing Fig.~\ref{conj-result} with Fig.~\ref{212p04fid}, increasing the bit flip noise probability reduces the learning speed of the quantum neural network. Moreover, as expected, the fidelity of the validation data set of $\text{QAE}_{\text{conj}}$ drops significantly by increasing noise probability. However, we will show in the next section using a modified decoder quantum circuit, the $\text{QAE}_{\text{conj}}$ will be trained even in the noise threshold $p=0.4$.

\section{ $\text{DQNN}_{\text{NISQ}}$-based QAE with a Conjugate Layer and a Modified Decoder}
\label{Modified decoder}
\begin{figure}[htb]
\centering
\includegraphics[width=\linewidth]{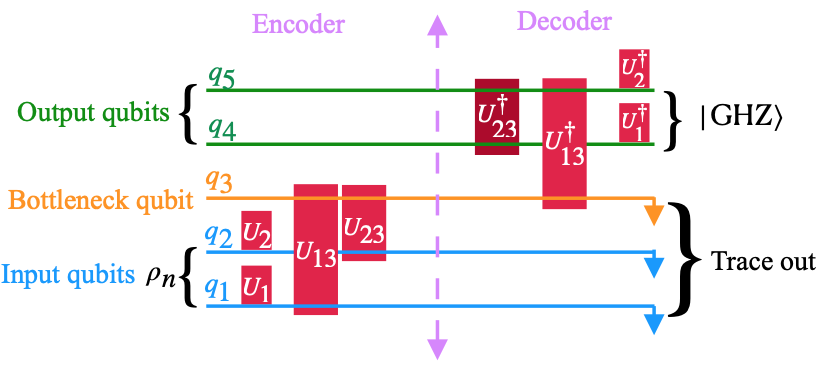}
\caption{Schematic configuration of a 2-1-2 architecture of $\text{DQNN}_{\text{NISQ}}$-based QAE with a conjugate layer and a modified decoder.}
\label{neda2}
\end{figure}
As mentioned in previous sections, despite the decrease in the learning parameters of QAE based on the $\text{DQNN}_{\text{NISQ}}$ with a conjugate layer, one cannot claim superiority of it compared to the $\text{DQNN}_{\text{NISQ}}$ without a conjugate layer. As illustrated in Fig.~\ref{QAE}, in the $\text{DQNN}_{\text{NISQ}}$, the qubits of each layer are not connected with two-qubit gates. Instead, each qubit of each layer is connected to all qubits of its adjacent layer. This particular model belongs to the category of a sparse QAE that has been previously discussed \cite{PhysRevLett.124.130502}.
Consider the reverse idea in the decoder part of QAE, i.e., the bottleneck qubit of the hidden layer connects to only one of the output qubits, and other output qubits connect to each other via two-qubit gates. Therefore, the bottleneck's information is indirectly transferred to all other output qubits. Figure ~\ref{neda2} shows the (2-1-2) QAE with the modified decoder and conjugate layer called $\text{QAE}_{\text{conj+mod dec}}$. In this new decoder scheme, the bottleneck qubit ($q_3$) connects to one of the output qubits ($q_5$) via the unitary operator $U_{13}^{\dagger}$. The intact output qubit ($q_4$) connects to this qubit through the two-qubit unitary operator $U_{23}^{\dagger}$. As a result, it gains to the bottleneck qubit information indirectly.

When the number of qubits of GHZ is more than two, the modified decoder can be more than one alternative. As depicted in Fig.~\ref{modified3-1-3}, depending on the number of qubits connected directly to the bottleneck qubits, three models exist for $\text{DQNN}_{\text{NISQ}}$ with a Modified Decoder. We will show the quantum network performs better as the number of qubits to connect to the hidden layer is decremented, i.e., model~3 represented in Fig.~\ref{modified3-1-3} (d).

\subsection{Simulation}

\begin{figure}[htb]
\centering
\includegraphics[width=\linewidth]{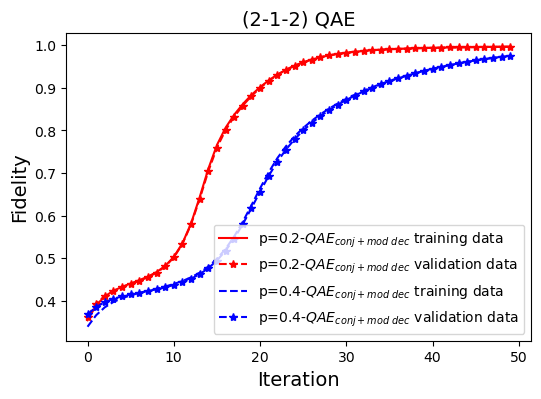}
\caption{(Color online) Comparison of the fidelity change during a (2-1-2) QAE training process based on $\text{QAE}_{\text{NISQ}}$ with a conjugate layer and modified decoder ($\text{QAE}_{\text{conj+mod dec}}$) while the bit-flip noise probabilities are  $p=0.2$ (solid red) and $p=0.4$ (dashed blue). The starred lines relate to the validation data applied to the learned quantum circuit during the iteration. The number of training data is 30. The optimization method is stochastic gradient descent (SGD) with a learning rate of 0.4.}
\label{neda-fid}
\end{figure}

\begin{figure}[htb]
\centering
\includegraphics[width=\linewidth]{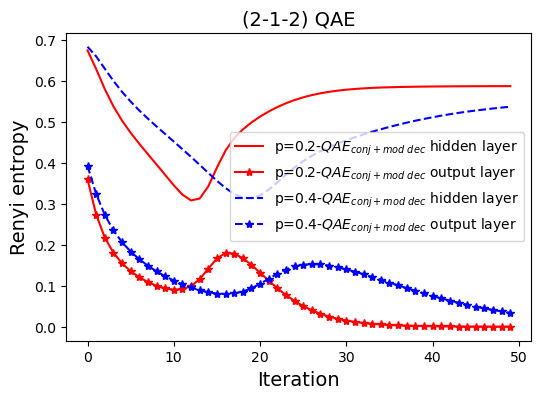}
\caption{(Color online) Comparison of the Renyi entropy change of output and hidden layer (bottleneck) during a (2-1-2) QAE training process based on $\text{QAE}_{\text{NISQ}}$ with the modified decoder ($\text{QAE}_{\text{conj+mod dec}}$). The bit-flip noise probabilities are $p=0.2$ (solid red) and $p=0.4$ (dashed blue). The starred lines relate to the output qubits Renyi entropy during the iteration. The number of training data is 30. The optimization method is stochastic gradient descent (SGD) with a learning rate of 0.4.}
\label{neda-reni}
\end{figure}

\begin{figure}[t]
\centering
\includegraphics[width=\linewidth]{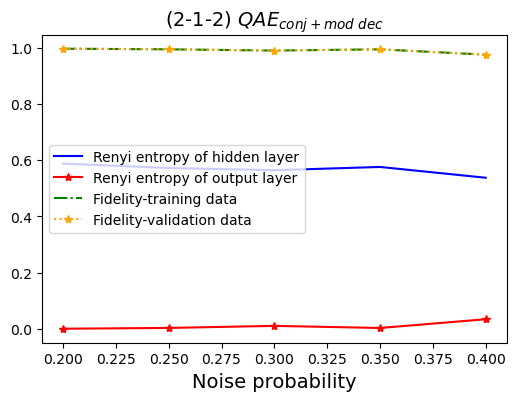}
\caption{ (Color online) Variation of the Fidelity and Renyi entropy of (2-1-2) $\text{QAE}_{\text{conj+mode~dec}}$ after 50 iterations with respect to different noise probability. The starred lines are related to the Renyi entropy of the output layer and the fidelity of validation data. The number of training data is 30. The optimization method is stochastic gradient descent (SGD) with a learning rate of 0.4.}
\label{noistol2}
\end{figure}

To compare the outperformance of the $\text{QAE}_{\text{conj+mod dec}}$ scheme with the results of section~\ref{DQNN+con}, the same training strategy, i.e., based on the average-based cost function, is regarded in this part.
Figure ~\ref{neda-fid} illustrates that using the modified decoder leads to the validation data coinciding with the training data for bit-flip noise probabilities $p=0.2$ and $p=0.4$, respectively. Moreover, the Renyi entropy of the output layers goes to zero. In contrast, in this new decoder scheme, the bottleneck qubit Renyis entropy is still correlated to the input qubits, as shown in Fig.~\ref{neda-reni}. 
Figure~\ref{noistol2} shows noise tolerance of $\text{QAE}_{\text{conj+mod dec}}$ after 50 iterations. Note that training and validation fidelities align for various noise levels.

\begin{figure}[htb]
\centering
\includegraphics[width=\linewidth]{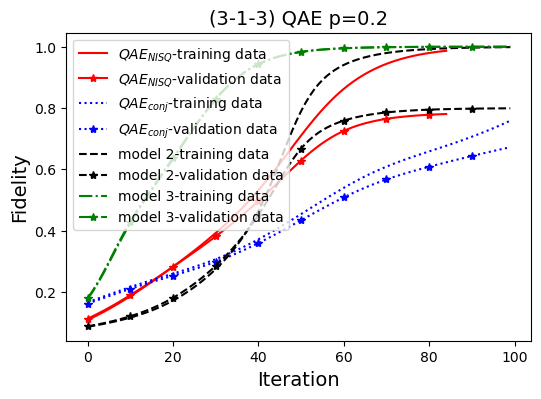}
\caption{(Color online) Comparison of the fidelity change of a (3-1-3) QAE training process based on $\text{DQNN}_{\text{NISQ}}$ equipped without a conjugate layer and a modified decoder $\text{QAE}_{\text{NISQ}}$ (solid red) and with a conjugate layer and without a modified decoder $\text{QAE}_{\text{conj}}$ (Model 0, dotted blue), with a conjugate layer and a modified decoder $\text{QAE}_{\text{conj+mod dec}}$, (Model~2 (dashed black), and Model~3 (dashdot green)).  The starred lines relate to the fidelity of the validation data applied to the learned quantum circuit during the iteration. The bit-flip noise probability is $p=0.2$.}
\label{fid23}
\end{figure}

\begin{figure}[htb]
\centering
\includegraphics[width=\linewidth]{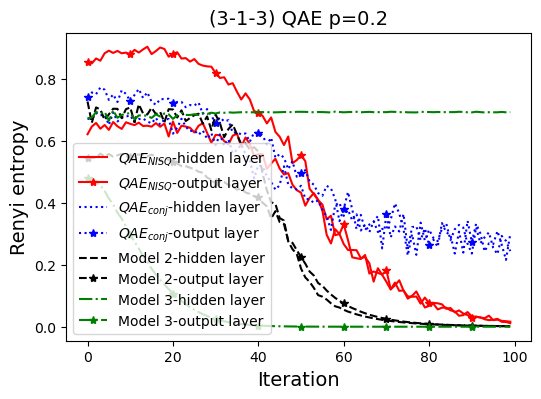}
\caption{(Color online) Comparison of the Renyi entropy change of outputs and hidden layer (bottleneck) during a (3-1-3) QAE  training process based on $\text{DQNN}_{\text{NISQ}}$ equipped without a conjugate layer and a modified decoder $\text{QAE}_{\text{NISQ}}$ (solid red) and with a conjugate layer and without a modified decoder $\text{QAE}_{\text{conj}}$ (Model 0, dotted blue), with a conjugate layer and a modified decoder $\text{QAE}_{\text{conj+mod dec}}$, (Model~2 (dashed black), and Model~3 (dashdot green)). The starred lines relate to the output qubits Renyi entropy during the iteration. The bit-flip noise probability is $p=0.2$.}
\label{ren23}
\end{figure}

\begin{figure*}[htp]
\centering
\includegraphics[width=\linewidth]{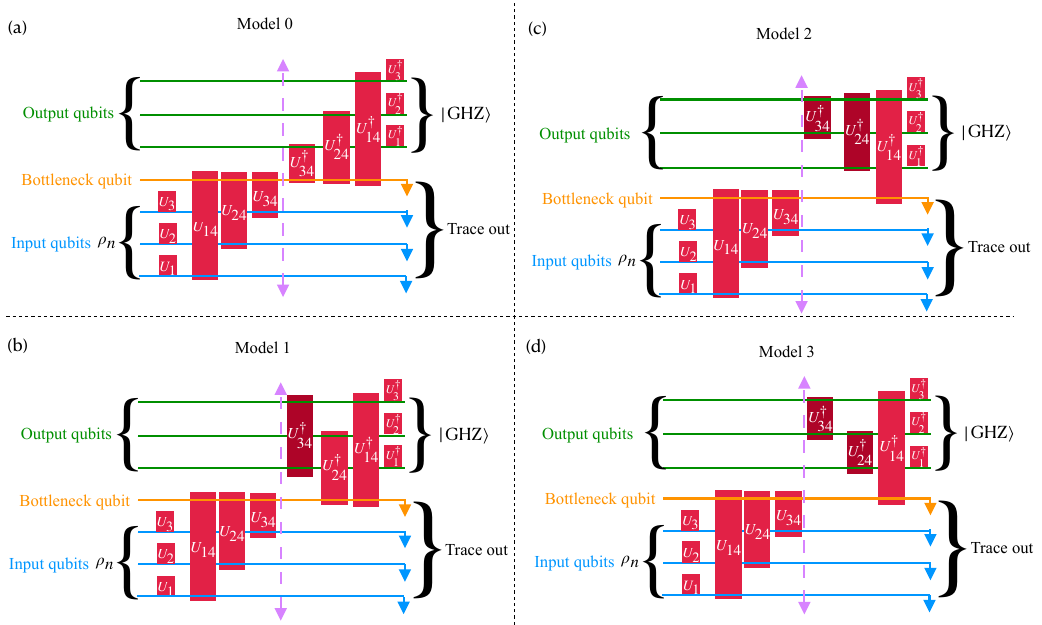}
\caption{(Color online) (a)  Schematic configuration of a 3-1-3 architecture of $\text{DQNN}_{\text{NISQ}}$-based QAE with a conjugate layer. (b-d) Schematic configuration of a 3-1-3 architecture of $\text{DQNN}_{\text{NISQ}}$-based QAE with a conjugate layer and a modified decoder (Model~1), (Model~2), and (Model~3).}
\label{modified3-1-3}
\end{figure*}

Model~0 ($\text{QAE}_{\text{conj}}$), Model~2, and Model~3 presented in Fig.~\ref{modified3-1-3}, corresponding to the (3-1-3) QAE, are compared with the training strategy based on the average-based cost function. Figures~\ref{fid23} and~\ref{ren23} illustrate this comparison for the bit-flip noise probability $p=0.2$ during the training via calculating the fidelity and Renyi entropy. Validation data confirms that model 3 is trainable, whereas models 1 and 2 are not. When comparing the variation in the fidelity of Model~0, Model~2, and Model~3 in Fig.~\ref{fid23}, it can be observed that as the connectivity of output layers becomes more indirect to the bottleneck's qubit, the QAE is trained better. For instance, the validation data of Model~2 reaches the same fidelity as the validation data of $\text{QAE}_{\text{NISQ}}$, while the learning parameters of the latter are two times the former.

\begin{figure}[t]
\centering
\includegraphics[width=\linewidth]{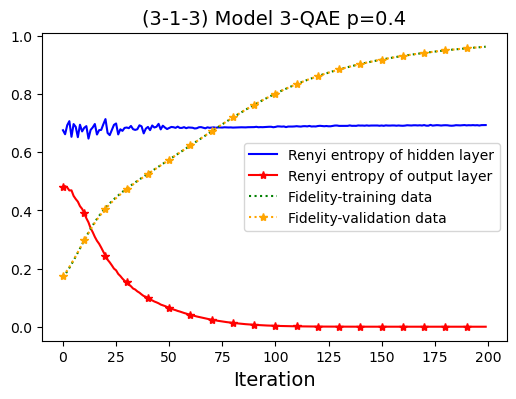}
\caption{ (Color online) The fidelity (dotted line) and Renyi entropy (solid line) change of a (3-1-3) QAE in the training process based on $\text{QAE}_{\text{conj+mod dec}}$ (Model~3).  The starred lines relate to the output qubits Renyi entropy (solid red) and the fidelity of the validation data (dotted orange) applied to the learned quantum circuit during the iteration. The bit-flip noise probability is $p=0.4$.}
\label{313p04m3}
\end{figure}

\begin{figure}[t]
\centering
\includegraphics[width=\linewidth]{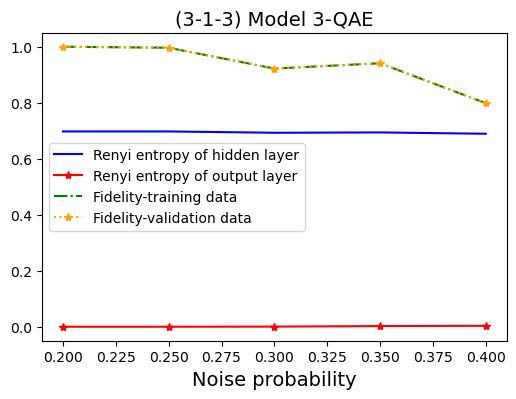}
\caption{ (Color online)  Variation of the Fidelity and Renyi entropy of (3-1-3) Model 3-QAE after 100 iterations with respect to different noise probability. The starred lines are related to the Renyi entropy of the output layer and the fidelity of validation data. The number of training data is 30. The optimization method is stochastic gradient descent (SGD) with a learning rate of 0.4.}
\label{noistol3}
\end{figure}

Figure~\ref{313p04m3} shows the trainability of the $\text{QAE}_{\text{conj+mod dec}}$ (Model~3) for the noise threshold $p=0.4$.  QAE denoised the GHZ state in this high noise threshold without utilizing the more hidden layers or brain box. Figure~\ref{noistol3} illustrates the noise tolerance of the $\text{QAE}_{\text{conj+mod dec}}$ (Model~3) after 100 iterations. It is evident from this figure that the training and validation fidelities consistently align for various noise levels.
Note that finding the best initialization of the learning parameters is always a crucial task in QML. Inspired by transfer learning \cite{5288526}, one can use the learning parameters of a $\text{DQNN}_{\text{NISQ}}$-based QAE to train a QAE equipped with a modified decoder and conjugate layer for denoising higher noise data.

\section{Conclusion}
\label{con}
This work proposes a new quantum network for training a QAE. Our proposed network is based on the dissipative quantum network utilizing conjugating layers. Accordingly, the learning parameters decrease. Modifying the quantum perceptron of the decoder of QAE, we demonstrate the noise threshold of QAE without brain boxing (deepening or stacking) is increased. The separability of the output GHZ (equivalently purity of GHZ) is investigated by calculating Renyi entropy. Throughout the learning process, we analyze the Renyi entropy of the hidden and output qubits in relation to other qubits. By keeping the hidden layer connected to the input layers during each iteration, our network can effectively denoise previously unseen noisy data with distinct underlying noise distribution. This is achieved by utilizing the learning parameters obtained from training data. We are confident that this approach will yield positive results and are enthusiastic about the potential impact it can have.

All the codes to simulate the results of this paper are based on the Pennylane and are found in the GitHub repository in \url{https://github.com/bathaee/QDNN-based-QAE}.

\clearpage
\bibliographystyle{apsrev4-1}
\bibliography{QAE-ref}

\end{document}